\documentclass[useAMS,usenatbib]{mn2e}
\usepackage{epsfig}
\usepackage{natbib}
\usepackage{color}
\usepackage{soul}
  \usepackage{aas_macros}

\newcommand{\kms}{km\,s$^{-1}$}
\newcommand{\vrad}{$v_{\rm rad}$}

\newcommand{\beii}{Be\,{\sc ii}}
\newcommand{\bevii}{$^{7}$Be}
\newcommand{\beviiii}{$^{7}$Be\,{\sc ii}}
\newcommand{\lii}{Li\,{\sc i}}

\newcommand{\liviiii}{$^{7}$Li\,{\sc ii}}
\newcommand{\livii}{$^{7}$Li}
\newcommand{\liviii}{$^{7}$Li\,{\sc i}}

\newcommand{\nai}{Na\,{\sc i}}
\newcommand{\cai}{Ca\,{\sc i}}
\newcommand{\caii}{Ca\,{\sc ii}}

\newcommand{\iiihe}{$^{3}$He}

\title[ \bevii\ in Nova Sgr2  ]{  Highly Enriched  $^{7}$Be in the ejecta  of Nova Sagittarii 2015  No.\,2 (V5668 Sgr) and the Galactic \livii\ origin }
\author[]{ Molaro, P.$^{1}$, Izzo, L.$^{2}$, Mason, E. $^{1}$, Bonifacio, P.$^{3}$, Della Valle, M.$^{4,5}$,  \thanks{E-mail:
molaro@inaf.oats.it (PM)}, 
  \thanks{ Based on observations collected at the European Souther Observatory, Chile.  Program
 ESO  DDT 294.D-5051.
 }\\
$^{1}$  INAF-Osservatorio Astronomico di Trieste, Via G.B. Tiepolo 11, I-34143 Trieste, Italy\\
$^{2}$  Instituto de Astrof\'isica de Andaluc\'ia (IAA-CSIC), Glorieta de la Astronom\'ia s/n, 18008 Granada, Spain \\
$^{3}$  GEPI, Observatoire de Paris, PSL Research Univ, CNRS, Univ Paris Diderot, Sorbonne Paris Cit\'e, Place J. Janssen, 92195, Meudon, France\\
$^{4}$ INAF-Osservatorio Astronomico di Napoli, Salita Moiariello, 16, I-80131 Napoli, Italy\\
$^{5}$ International Center for Relativistic Astrophysics, Piazza della 
Repubblica 10, I-65122 Pescara, Italy
 }
\begin{document}

\date{Accepted.... Received 2016-07-31}

\pagerange{\pageref{firstpage}--\pageref{lastpage}} \pubyear{2002}

\maketitle

\label{firstpage}

\begin{abstract}

We report 
 on  the  evidence  of highly blue-shifted resonance lines of the singly ionised 
isotope of \beviiii ~  in high resolution   UVES  spectra of  Nova Sagittarii 2015 No.\,2 (V5668 Sgr).   
The   resonance doublet lines \beviiii ~  at $\lambda\lambda$313.0583,\,313.1228  nm are  clearly detected in several non saturated and partially resolved high velocity components during the evolution of the outburst. The  total absorption identified with  \beii ~  has  an equivalent width  much larger   than all other elements and  comparable  to hydrogen.   We  estimate  an     atomic fraction   $N(\mbox{\bevii})/N(\mbox{Ca})$   $\approx$ 53-69 from  unsaturated and resolved absorption components. The   detection  of $^{7}$Be in several high velocity components  shows  that $^{7}$Be has
 been freshly created in a  thermonuclear runaway via the reaction   $^{3}\mbox{He}(\alpha,\gamma)^{7}\mbox{Be}$ during the Nova 
 explosion, as postulated by \citet{arn75},  however in  much larger amounts  than predicted  by  current models. 
  \beviiii ~     decays to   \liviiii\    with a half-life of 53.22 days,   comparable to the  temporal span covered by the observations.  The non detection of   \liviii\  requires that   \livii\  remains  ionised throughout our 
observations.
 The massive  \beii~  ejecta result into a  \livii\    production that is  $\approx$  4.7-4.9 dex  above  the meteoritic abundance. If  such a high   production is common  even in a small fraction ($\approx$5\%) of  Novae, they  can make all the {\it stellar}  \livii ~   of the Milky Way.

\end{abstract}


\section{Introduction}

\livii  ~ is a  unique element that shows a large  variety of  production processes.
  These  include primordial nucleosynthesis,  spallation processes by  high energy  
 cosmic rays in the interstellar medium,  
 stellar  flares in  low mass
  stars, Cameron-Fowler mechanism in  Asymptotic Giant Branch (AGB) stars and Novae, and neutrino induced   nucleosynthesis in SNae explosions. Observations show that \livii ~ has a constant abundance among metal-poor stars and begins to rise at [Fe/H] $\approx$ -1 to reach the meteoritic value at solar metallicities \citep{reb88} requiring   a net \livii~ production   \citep{rom99}.  
The rate of the Li increase    favours     AGB stars  and Novae  as the  most significant  $\it stellar$ sources.  
Although \livii ~ has been observed in AGB stars the 
 observational evidence for Novae   has  only recently been found by \citet{izz15}   with  the first detection  of the 
\liviii  ~ $\lambda\lambda$6708 line in 
the  spectra of  
Nova Centauri 2013 (V1369 Cen)     and by 
  \citet{taj15}   with   the first detection of $^{7}$Be    in 
  the post-outburst spectra of the classical 
  Nova Delphini 2013 (V339 Del).

 Here, we report a study of  the  \beii  ~ by means of   UVES observations of  Nova Sagitarii  2015 No.\,2  (V5668 Sgr).
A spectrum from the  High Dispersion Spectrograph of the Subaru Telescope  taken  at day  63 after maximum has been   discussed 
 by \citet{taj16}   who reported   the    presence of \beviiii ~  in  this Nova,  and also in V2944 Oph.
 
\section{Observations }

 \subsection{Evidence for \beviiii}

 Nova Sagittarii 2015 No.\,2  was discovered  by \citet{sea15}   on  15 March  2015  and reached the first maximum   on 21  March  at 04h 04m {\sc UT} with a magnitude of  $V = 4.3$. 
The Nova  re-brightened several times and   remained    bright    for about 80 days before  declining  due to  dust formation.
 Soon after the discovery we started a  DDT program with the UVES spectrograph at the ESO-VLT.  Several UVES spectra    were obtained
at  +58, 63, 69, 73,  82 and 89
days from maximum as reported in Table \ref{tab}. The settings with central  wavelength of 346\,nm (range 305-388\,nm), 437\,nm (375-499\,nm), 564\,nm (460-665\,nm) and 760\,nm (570-946\,nm), were used, thus covering the full optical range from the atmospheric cutoff  to the red edge of 946.0 nm with small gaps of $\approx$ 10 nm around the red central wavelengths. The resolving power   was $R= \lambda /\delta \lambda  \approx 80,000$ for the blue arm and $\approx 120,000$ for the red arm.  Overlapping spectra have been combined for each epoch to maximise the signal-to-noise.

\begin{figure}
\centering
\includegraphics[width=8.cm,angle=0]{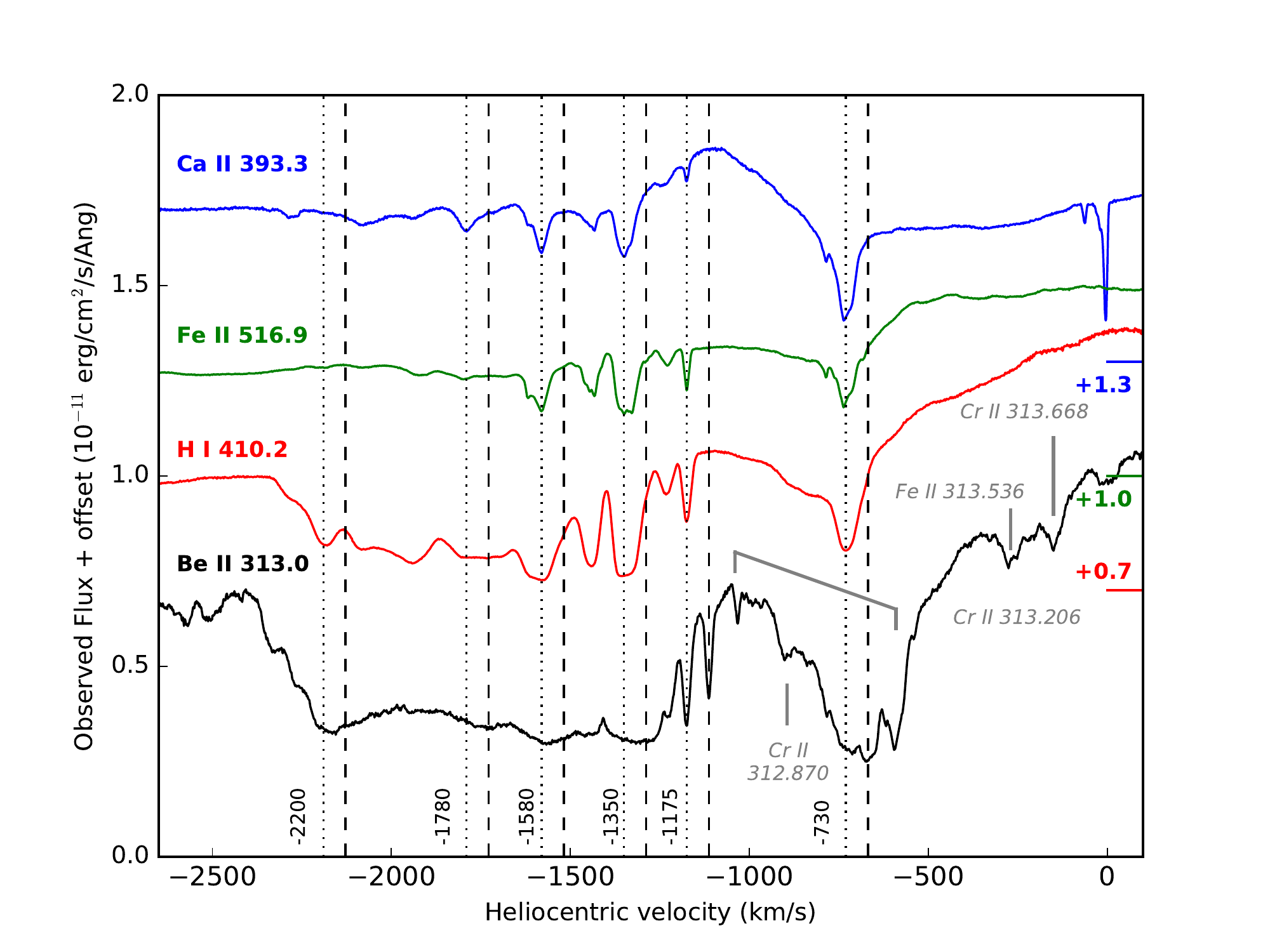}
 \caption{ Nova  spectrum at day 58.  The figure  displays  
 the spectrum in the vicinity of Fe\,{\sc i}\ $\lambda\lambda$519.6  (green line)  H\,{\sc $\gamma$}  (red), Ca\,{\sc ii}\,K (blue)
 and   $^{7}$Be\,{\sc ii} (black) lines  plotted
 on the velocity   scale.  The fluxes are scaled to provide roughly the same intensity shortwards the absorption.  The velocity scale is adjusted to  the $^{7}$Be\,{\sc
 ii}  $\lambda\lambda$313.0583 nm line. The position of the main absorption components is shown with vertical  dotted lines. The corresponding position of the expected  $^{7}$Be\,{\sc
 ii}  $\lambda\lambda$313.1228 nm line is shown by vertical   dashed line.
  }
 \label{fig1}
\end{figure}

\begin{figure}
\centering
 
\includegraphics[width=8.5cm,angle=0]{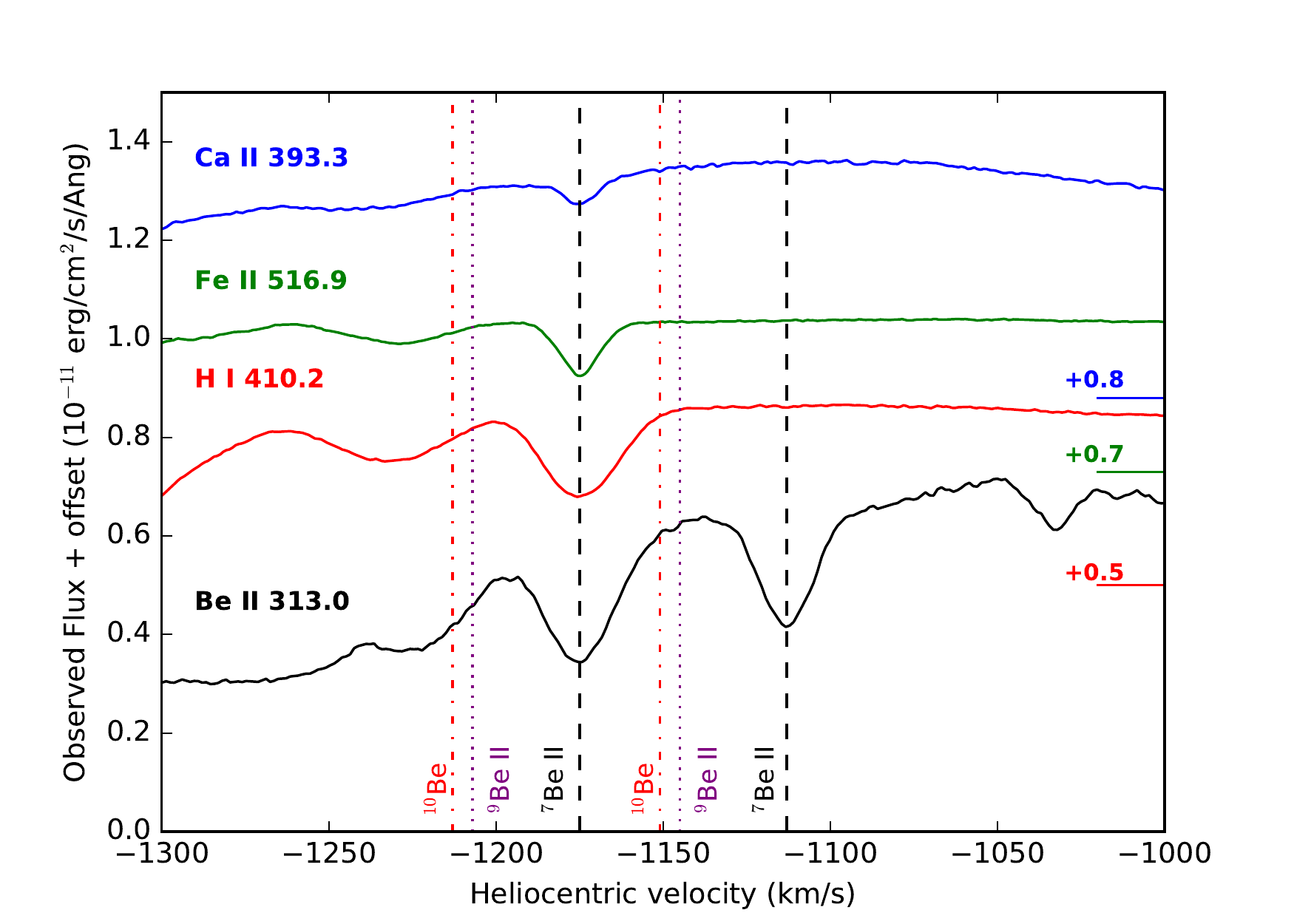}
 \caption{  Zoom  of   Fig. \ref{fig1} around the $-1175$ \kms component. The \beviiii ~ $\lambda\lambda$313.0583,\,313.1228 nm are shown with a dashed line.  The  positions of the expected  $^{9}$Be\,{\sc
 ii}     $\lambda\lambda$313.0442,\,313.1067  nm   and  $^{10}$Be\,{\sc ii}  $\lambda\lambda$313.0484,\,313.1129  nm are also shown.}
 \label{fig2}
\end{figure}

At early epochs the spectra of the Nova   show  several  broad emission lines of neutral hydrogen 
 and other permitted transitions of neutral
or singly ionised species   often  accompanied by sharp and
blue-shifted multiple absorption components reaching blue edge    velocities of $\approx$ -2300   \kms.
 Fig \ref{fig1} displays portions of the Nova spectrum of day 58 
in the proximity of   H\,{\sc $\gamma$}, Ca\,{\sc ii}\,K , Fe\,{\sc ii} $\lambda\lambda$ 519.0 and  Be\,{\sc ii} $\lambda\lambda$313.0 nm lines. 
 These  lines   show several  absorption components  at  heliocentric  \vrad $\sim$ $-730,  -1175, -1350,  -1580,  -1780$ and  $-2200$ \kms  with the more prominent ones marked  with vertical black dotted lines in the figure.
 Ca\,{\sc ii}\,K 
shows also   narrow  absorption components  at  $-4.3$ and $-62.9$ \kms\ velocities   caused by intervening Galactic interstellar  medium.
At the wavelength of the Be\,{\sc
 ii}  $\lambda\lambda$313.1 nm doublet  there is P-Cygni profile with a huge blue-shifted  absorption that  \citet{taj16} identified as    $^{7}$Be\,{\sc}.  
In Fig \ref{fig2} we   zoomed  the   component  at $-1175$  \kms which is seen only at this epoch and provides   a robust identification.   The two  sharp absorption components   (FWHM $\approx$ 0.19 \AA ) are separated by  0.654 \AA ~ which corresponds  precisely to the  separation   of the    $^{7}$Be\,{\sc ii} resonance doublet of $\lambda\lambda$313.0583,\,313.1228 nm.  
  Moreover,    the component is  perfectly aligned in radial velocity with   the other  species   providing  evidence that it is \bevii\,{\sc ii} and not   $^{9}$Be\,{\sc ii}  which has an isotopic shift of $-15.4$  \kms. 
  The dips of each line of the $^{7}$Be\,{\sc ii}  doublet can be  identified  also at $-730$ \kms   in Fig \ref{fig1} , but become hard to  see in the  flat bottoms of   the other components.  At  velocities   $\le$  $-1600$ \kms ~  the absorption profile      ascribable  to Be follows mainly  the hydrogen lines rather than the metallic lines which are very weak or absent.   If based only  on this spectrum the identification   of this part of absorption with $^{7}$Be\,{\sc
 ii} would therefore be   controversial.

\begin{table}
\caption{Journal of the observations.  \label{tab}}
\begin{center}
\scriptsize
\begin{tabular}{lrrrlrrr}
\hline
\hline
\multicolumn{1}{c}{Date} & 
\multicolumn{1}{c}{{UT}} & 
\multicolumn{1}{c}{{MJD}} & 
\multicolumn{1}{c}{{Day}} & 
\multicolumn{1}{c}{346}& 
\multicolumn{1}{c}{437}& 
\multicolumn{1}{c}{564}&
\multicolumn{1}{c}{760}\\
\multicolumn{1}{c}{2015}& 
\multicolumn{1}{c}{h m} & 
\multicolumn{1}{c}{57000} & 
\multicolumn{1}{c}{a.m.}&
\multicolumn{1}{c}{s} & 
\multicolumn{1}{c}{s} &
\multicolumn{1}{c}{s} &
\multicolumn{1}{c}{s}  \\
\hline
\hline
 05-19            & 08 49 & 161.36&  58.7 &  400     &2x150  &2x100 &2x150\\
 05-24            & 05 49 &  166.24 & 63.6  & 2x400 &2x150   &4x100& 2x150 \\
 05-30            & 03 20 &  172.14 & 69.5  &  400 &  100 &9x7&3x10\\
 06-03            & 05 18 & 176.22& 73.6  &  400 & 100 &9x7&3x7\\
 06-12            & 05 25 & 185.23 &   82.6    &400 & 60   &9x7&2x10\\
06-19           & 01 29 &   192.06 &89.4  &  2x602& 2x60  &22x15 &3x15\\
\hline
\end{tabular}
\end{center}

\end{table}

\begin{figure}
\centering
 \includegraphics[width=8.5cm,angle=0]{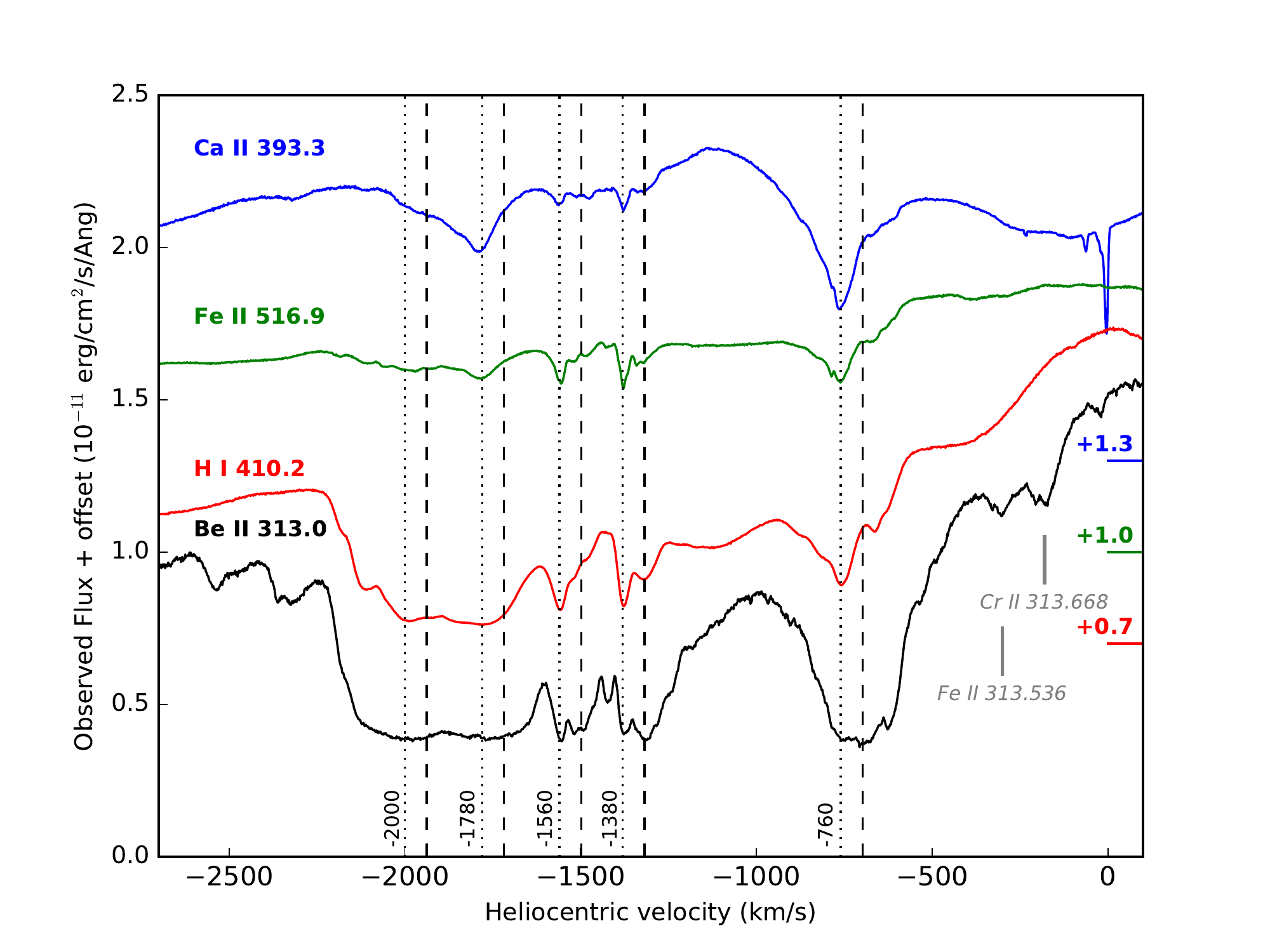}
 \caption{  Same as  Fig \ref{fig1} with the Nova spectrum  of   day 63.  Legend as in  Fig. 1.  Note the presence of the  partially resolved components of the $^{7}$Be\,{\sc
 ii}  $\lambda\lambda$313.1228 nm line for the components at velocities $\approx$ $-1380$ and  $-1560$ \kms.
   }
 \label{fig3}
\end{figure}

\begin{figure}
\centering
 \includegraphics[width=8.5cm,angle=0]{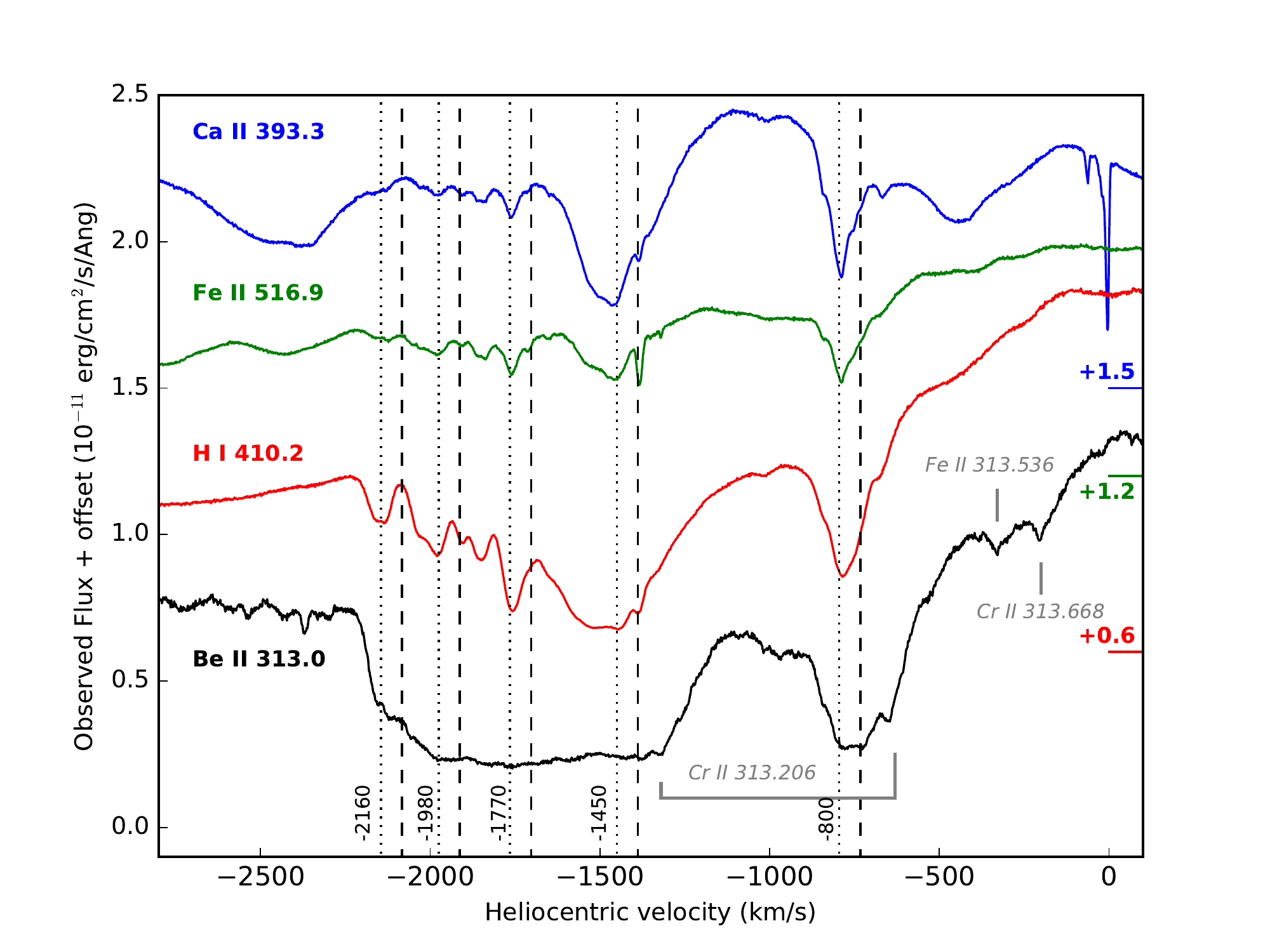}
 \caption{  Same as Fig \ref{fig1} with the Nova spectrum at  day 69.  This observation is  close to the one discussed in \citet{taj16}.
   }
 \label{fig4}
\end{figure}

\begin{figure}
\centering
 \includegraphics[width=8.5cm,angle=0]{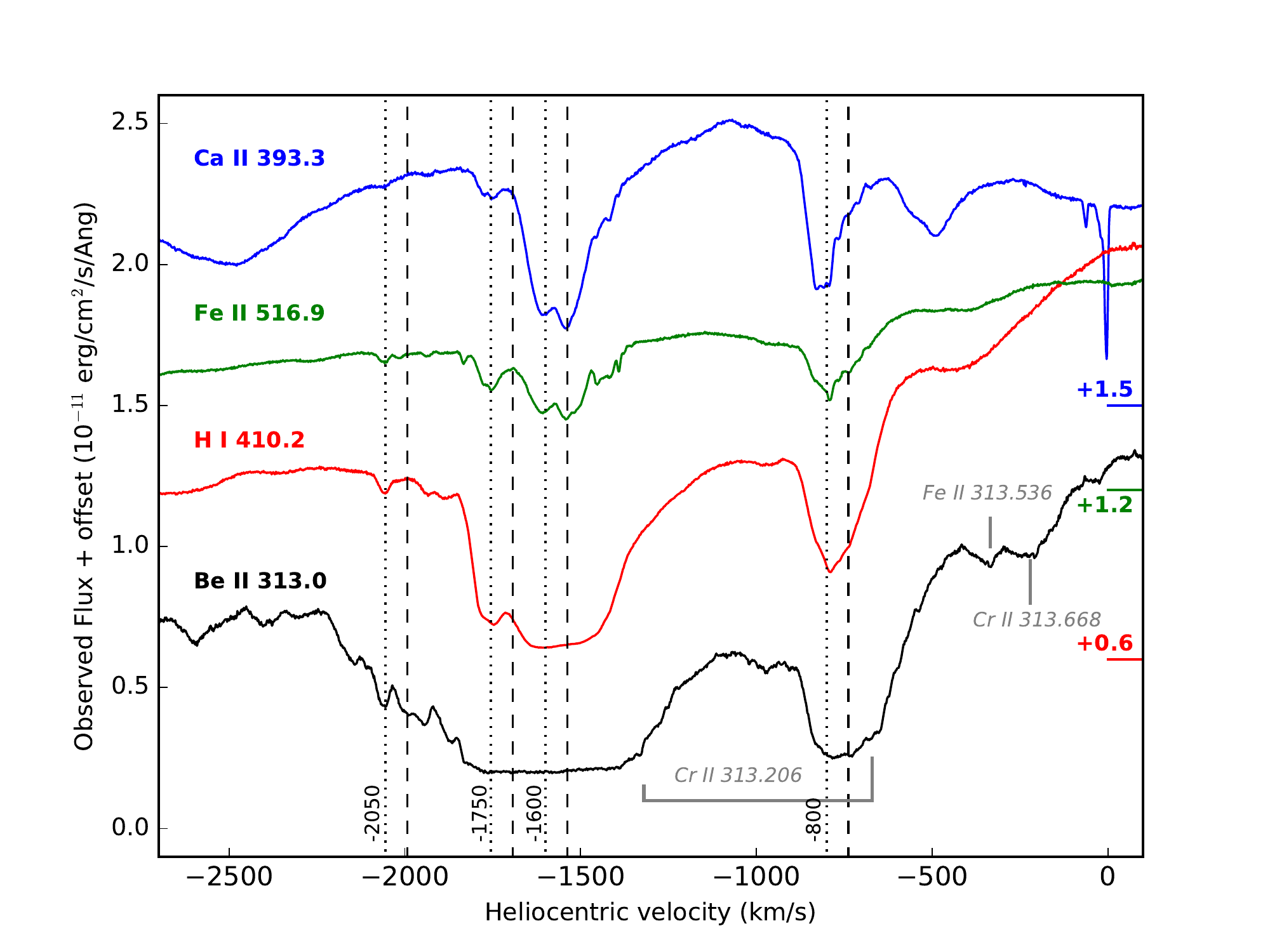}
 \caption{  Same as  Fig \ref{fig1} with the Nova spectrum
   at day 73.    The $^{7}$Be\,{\sc
 ii}  313.1228 nm line  is   partially resolved  in the  component at velocities $\approx$ $-2050$ \kms. 
   }
 \label{fig5}
\end{figure}

\begin{figure}
\centering
 \includegraphics[width=8.5cm,angle=0]{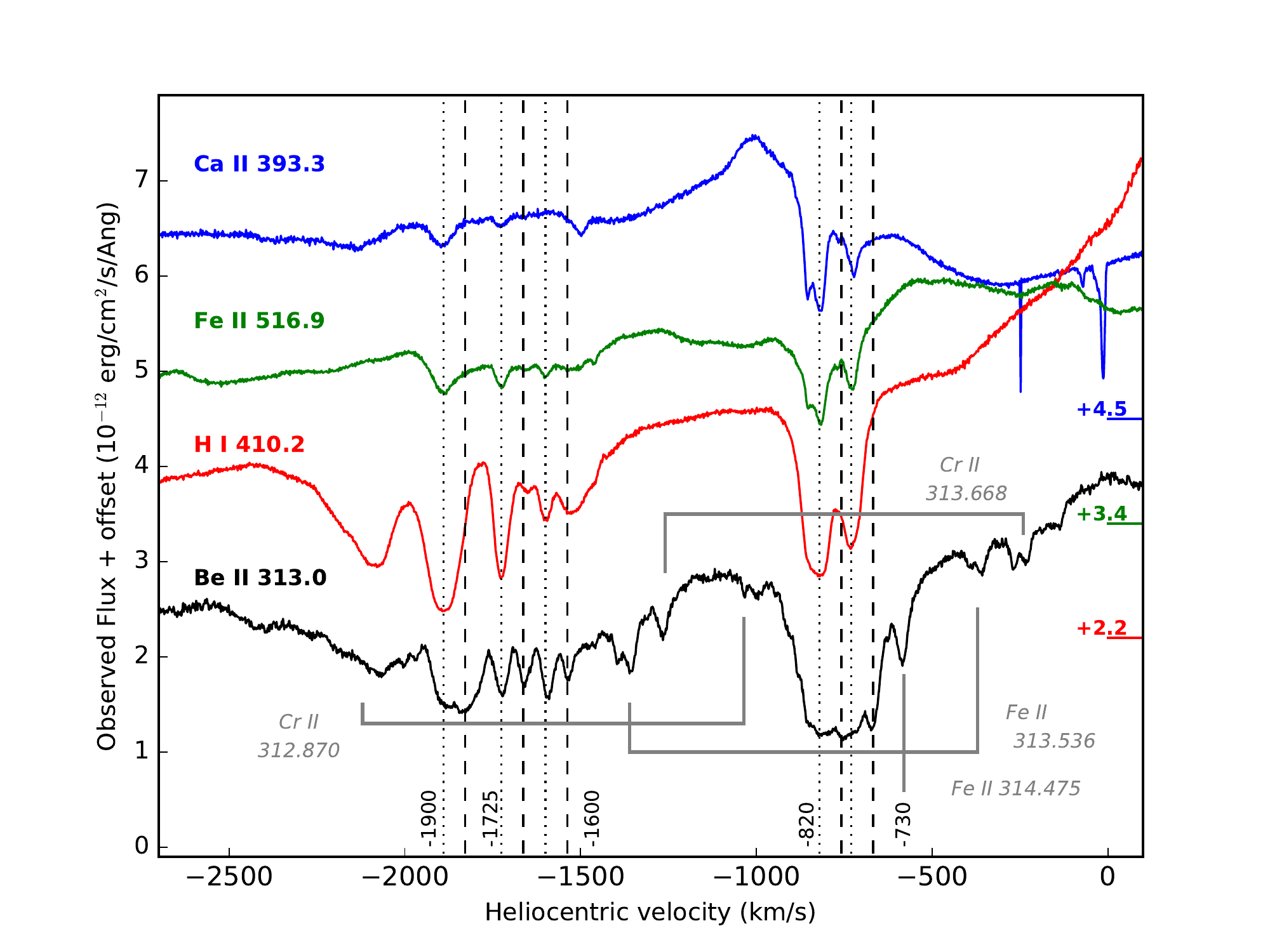}
 \caption{  Same as  Fig \ref{fig1} with the Nova spectrum  at day 82. The $^{7}$Be\,{\sc
 ii}  313.1228 nm line  is   partially resolved  in the two components which break up at velocities $\approx$ $-820$ \kms and  fully resolved in the components  at velocities  at $\approx$ $-1600$  and $-1725$ \kms which were previously strongly saturated. Note the components at $-1900$ \kms of the CrII and FeII which are now visible due to the \bevii ~ weakening.
   }
 \label{fig6}
\end{figure}

\begin{figure}
\centering
 \includegraphics[width=8.5cm,angle=0]{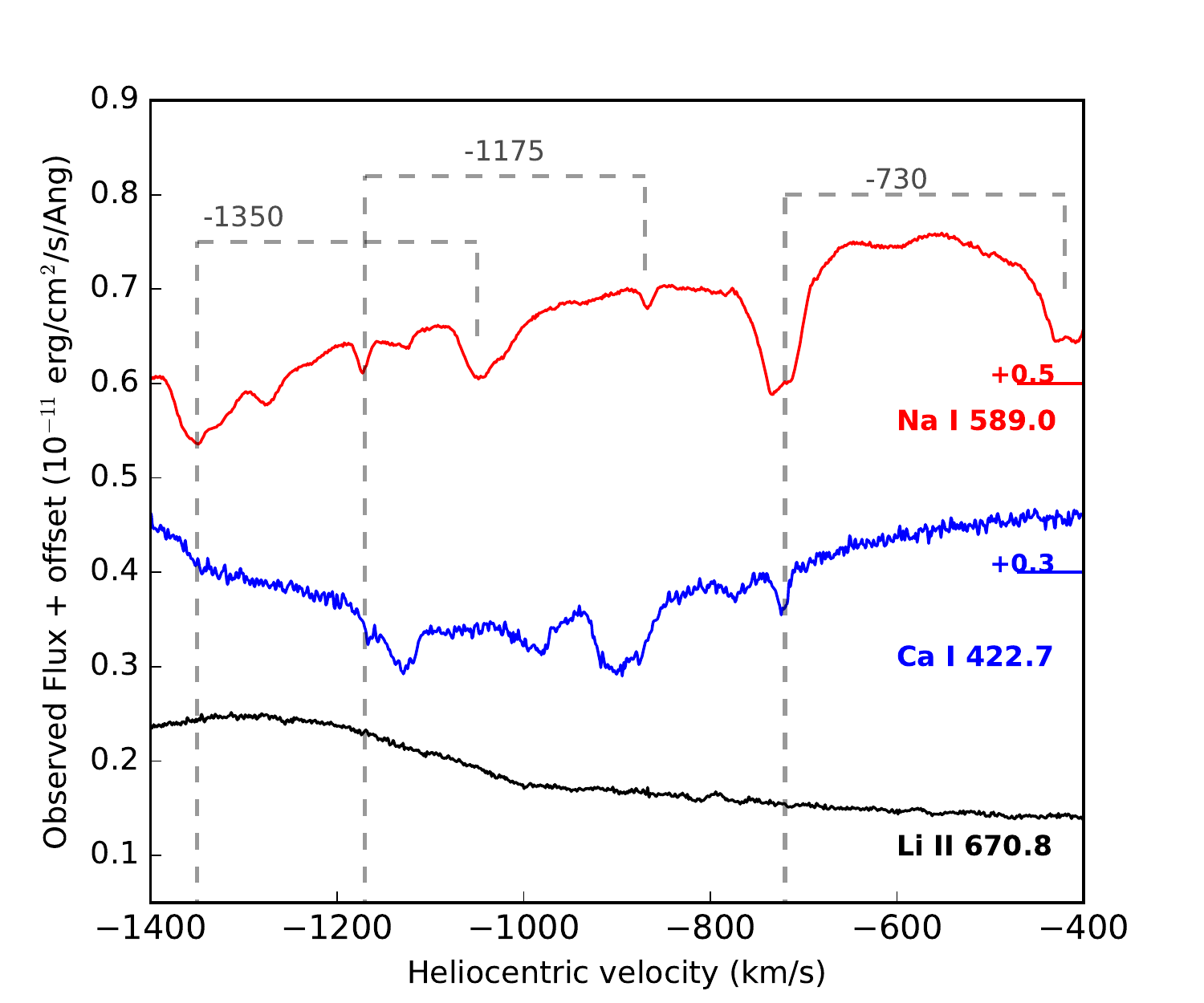}
 \caption{  
 Nova spectrum at day 58 in the vicinity of \liviii ~ $\lambda\lambda$670.8 nm   (black line),   Na\,{\sc i}\ D doublet (red) and Ca\,{\sc i}\,$\lambda\lambda$ 422.7 nm  (blue).   }
 \label{fig7}
\end{figure}

 The bottoms of the  strong lines  in Fig \ref{fig1} are  totally  flat   suggesting that the absorption
is saturated  but with the  absorbing
material only  partially covering the background light source.  The Balmer lines also show flat  bottoms.  At this epoch   in correspondence of the  $^{7}$Be\,{\sc
 ii} the  intensities  are $\sim$50 \% but  the  intensity value  varies with   the day and the geometry of the outburst.

The \beviiii\ absorption may be contaminated by the presence of  
other Fe-peak elements. Evidence is found  for the presence of 
 Cr\,{\sc ii}\,(5)  multiplet. The   Cr\,{\sc ii}\ $\lambda\lambda$ 313.2056  nm  line is observed resolved both in the $\sim -730$, $\sim -1175$ \kms components.      The  Cr\,{\sc ii}\,(5)$\lambda\lambda$ 313.6680  and 312.8699  nm and the  Fe\,{\sc ii}\,(82) $\lambda\lambda$313.5360  nm  lines are now seen     at $-730$ \kms.  The Cr\,{\sc ii}\,  $\lambda\lambda$ 312.4978 nm line of the same multiplet (5) and with  comparable intensity should therefore be present and contribute to the main absorption. Other lines  show up on  day 82 and are listed in Table \ref{tab2}.

Fig \ref{fig3} displays portions of the Nova spectrum obtained on day 63.  The components at velocities $\approx$ $-1380$ and  $-1560$ \kms become  sharper and it is possible to identify  the  partially resolved component of the $^{7}$Be\,{\sc
 ii}  $\lambda\lambda$313.1228 nm line  shown in the figure by vertical blue dashed lines. In addition,  some absorption consistent with its presence can be observed in the $-760$ \kms component as well as in all other lines. It is quite remarkable that while H$\gamma$  shows some structure, \bevii ~ does not,  and this is  likely due to the presence of the doublet lines filling the inter-component velocity-space. 
 Fig \ref{fig4} displays  the  spectrum  obtained  on    day 69  which is very close to  the  spectrum analysed by \citet{taj16}. As it can be seen   
  the  narrow component  identified by  \citet{taj16}  as  $^{7}$Be\,{\sc  ii} $\lambda\lambda$ 313.1228 nm is  blended with  the $-1450$ \kms  component of the  Cr\,{\sc
 ii}  $\lambda\lambda$313.2058 nm.
Fig \ref{fig5} displays portions of the spectrum obtained on day 73.  At this  epoch   the  high velocity  components at   $\approx$ $-2000$ \kms weaken considerably in   $^{7}$Be\,{\sc
 ii}   revealing the presence of the  $^{7}$Be\,{\sc
 ii} $\lambda\lambda$ 313.1228 nm line in the  components   marked with a  dashed blue line in the figure.   
 To note that  at this day the  $^{7}$Be\,{\sc
  ii}    absorption spans   wider velocities than H$\gamma$.
Fig \ref{fig6} displays portions of the  spectrum obtained on day 82. At this epoch the high velocity components of $^{7}$Be\,{\sc
 ii}   weaken considerably   revealing  few resolved components  previously  hidden  inside the absorption.
The $^{7}$Be\,{\sc
 ii}  $\lambda\lambda$313.1228 nm line  is  partially  resolved  in the two components which break up at velocities $\approx$ $-820$ \kms and is fully resolved in the components  at velocities  between $\approx$ $-1600$  and $-1725$ \kms which were previously strongly saturated.
 After few days from this observation all the metallic components disappeared and therefore later observations are not considered here.
 Due to the weakening of the  $^{7}$Be\,{\sc
 ii}  on day 82 also the metallic contaminants due to  iron-peak elements appear very clearly.   The  Cr\,{\sc ii}\,(5) $\lambda\lambda$313.6680   nm and the  Fe\,{\sc ii}\,(82) $\lambda\lambda$313.5360  and 314.4751 nm  lines are now seen   at $-1900$ \kms  and 
   should have been  present also in the previous epochs but completely obscured by the  strong  $^{7}$Be\,{\sc
 ii}  absorption.  The  Cr\,{\sc ii}\,(5) $\lambda\lambda$313.6680  and 312.8699  nm and the  Fe\,{\sc ii}\,(82) $\lambda\lambda$313.5360  nm  lines are now seen  at $-820$ \kms.  Scaling with the relative strengths of the other iron-peak elements 
    we estimate that the combined contributions of these contaminants 
are $\approx$ 3.5  \%\  of the total equivalent widths of the \beviiii\ absorption.

\subsection{\bevii\ abundance}

 The abundance of \bevii\  can be  estimated by comparing   the equivalent widths 
$W(\mbox{\beviiii})$ and  $W(\mbox{\caii\,K})$ for 
 unsaturated and   resolved lines,  assuming   that  the covering factor of the absorbing expanding shell is constant with  wavelength.  Ca is not a Nova product and can be taken as a reference element. The more suitable  components are     the   features of \beviiii\ and  the \caii\,K lines observed at $-1500$ \kms\  on  day 82.  Though, we are  aware that  
these abundances       do not necessarily represent the abundances in the whole materials ejected.
 
Following  \citet{spi98}   and \citet{taj15} the ratio of column  densities, $N$, can be written as

\bigskip

\begin{eqnarray}
  \frac{N(\mbox{\beviiii})}{N(\mbox{\caii})}
  & = &   2.164   \times \frac{W(\mbox{\beviiii\,, Doublet})}{W(\mbox{\caii\,, K})} 
\end{eqnarray}

\bigskip

For the components at $-1500$ \kms\ on day 82 we measured  $W(\mbox{\beviiii})$ = (0.095 +0.060) = 0.155 \AA~ and the $W(\mbox{\caii\,K})$ = 0.019 \AA~  which provide
a column density ratio  of  $N(\mbox{\beviiii})/N(\mbox{\caii})$ =
  $ 17.7$.   Assuming  that most of \bevii\ and  Ca are  in the singly ionised state as discussed in \citet{taj16}  these  are also the relative elemental abundances.
The  presence of \nai\  in the blue-shifted absorption line systems along with the presence of \cai\, on day 58\relax\ , as shown below,  and the absence of doubly ionised iron-peak elements 
 support this assumption. 
 Since our measurement refers  to day  82   after  maximum  and $^{7}$Be  decays to  $^{7}$Li    via  K-electron capture   with a half-life of 53.22 days,  the amount of \bevii\ freshly produced  by the Nova   should have been 
$\approx$   3  times larger which gives  an atomic fraction   $N(\mbox{\bevii})/N(\mbox{Ca})$  of $\approx$ 53.  
We can determine the  \bevii ~ abundance also for the component at $-1175$ \kms\ on day 58, which is fully resolved but slightly saturated. In this case  we have    $W(\mbox{\beviiii})$ = (0.089 +0.073) = 0.162 \AA~ and the $W(\mbox{\caii\,K})$ = 0.011 \AA~  which provides  a   $N(\mbox{\beviiii})/N(\mbox{\caii})$ = 31.9. Considering that  on day 58 the original value should have been  a factor 2.15 larger,  we obtain   an original  atomic fraction   $N(\mbox{\bevii})/N(\mbox{Ca})$  of $\approx$ 69, which is quite consistent with the former value considering the uncertainties involved. \citet{taj16} derived a N(\beviiii)/N(\caii) = 8.1 $\pm$ 2.0 in the component at $-$786 \kms on day 63 without considering \beviiii~ decay. Since this component is saturated, the derived abundance is a lower limit and therefore the  two measurements  are consistent with each other.

\subsection{ Nova \bevii\ production}

 Thermonuclear production of \bevii\    during the Nova explosions of hydrogen-rich layers containing some \iiihe\  has been proposed by \citet{arn75} and \citet{sta78}. Peak temperatures of ~ 150 million K are reached in the burning regions and 
  \bevii\  is readily formed   from the \iiihe\  
coming from the   companion star via the reaction
 \iiihe\ ($\alpha$, $\gamma$)\bevii ~    \citep{her96}. In this hot environment \bevii\    can be also destroyed and it 
 needs to be carried to cooler regions by convection on a time scale  shorter than the destruction time scale as in the Cameron-Fowler mechanism \citep{cam71}.
 The cooler regions    are subsequently ejected and observed in absorption in the Nova outburst.  
 Carbon and oxygen (CO) Novae  destroy   less   \iiihe\  with respect to oxygen and neon (ONe ) Novae, and therefore   CO Novae    have higher    \bevii\   yields \citep{jos98}.

 The detection of \bevii\ in the post-outburst spectra of Nova Sagittarii 2015 shows that 
 thermonuclear  production of \bevii\ is effectively taking place.  The fact that \bevii\ is detected  at all velocities   implies  that 
 all the absorption components are made of    ejecta
 which have experienced thermonuclear runaway nucleosynthesis.
 However, the observed yields are larger by about one order of magnitude than predicted by the  models of    \citet{ jos98}and even more if compared  with the models of \citet{bof93}.
 The  number of freshly produced \bevii\ atoms  in Nova ejecta is  necessarily lower than   that of  \iiihe\ atoms in the accreted gas from the donor star or produced in situ  as a result of the so called  \iiihe\  bump \citep{den13}. This implies that 
 the fraction of  (\iiihe/H)   should be greater than $ 10^{-4}$.   

\begin{table}
\begin{center}
\caption{Contaminants  from Fe-peak elements in the range of the \beviiii\ doublet measured  in the spectrum of  day 82. \label{tab2}}
\begin{tabular}{lcrccc}
\hline
\hline
\multicolumn{1}{c}{Lines}&
\multicolumn{1}{c}{$\lambda_{\rm lab}$(\AA)}&
\multicolumn{1}{c}{$\log gf$}&
\multicolumn{1}{c}{$W$ (m\AA)}& 
\multicolumn{1}{c}{$W$  (m\AA)}\\
\multicolumn{1}{c}{}&
\multicolumn{1}{c}{}&
\multicolumn{1}{c}{}&
\multicolumn{1}{c}{$-820$}& 
\multicolumn{1}{c}{$-1900$ }\\[2pt] \hline
Cr\,{\sc ii}\,(5)  & 3120.3691& $ 0.120$ & 16: $\pm$ 8  & -  &  \\
Cr\,{\sc ii}\,(5)  & 3124.973 & $ -0.018$ & b & b  &  \\
Cr\,{\sc ii}\,(5)  & 3128.700 & $-0.320$ &    25 $\pm$ 7    & b &  \\
Cr\,{\sc ii}\,(5)  & 3132.053 & $0.079$ &  b  & b    &   \\
Fe\,{\sc ii}\,(82)  & 3135.360 & $ -1.130$ &   40 $\pm$ 7 & 104 $\pm$ 10&  \\
Cr\,{\sc ii}\,(5)  & 3136.681 & $-0.250$ &     45 $\pm$ 8  & 51 $\pm$ 15&  \\
Fe\,{\sc ii}\,(82)  & 3144.751 & $ -1.740$ &  b &  83 $\pm$ 10 &  \\
\hline
\end{tabular}
\end{center}
\end{table}

 \subsection{ Nova  \livii\  production}

The \bevii\    decays to   \livii\ ~    with a half-life of 53.22 days  which is comparable with our temporal span. However, 
 we do not detect    counterparts of blue-shifted absorption line systems of
 the \liviii\, $\lambda\lambda$670.8 nm     in spite of the high signal-to-noise ratios 
 of our  spectra.  
 Figure \ref{fig7}  displays  
 the spectrum on day 58\relax\ in the vicinity of \liviii\,  $\lambda\lambda$670.8 nm,   Na\,{\sc i}\ D doublet  and Ca\,{\sc i}\, $\lambda\lambda$422.7 nm  
  lines. The other epochs are similar with the only difference that the trace  Ca\,{\sc i}\, disappears.
  We note however  that  while   Ca\,{\sc i}\, is non present the  \nai\,D lines    are relatively strong      and the complete absence of \livii ~ is rather puzzling.    
 It is interesting to report that
 these lines  have  been detected 
 in the first three weeks spectra  of  Nova Centauri 2013    \citep{izz15} and 
  that  \citet{izz16}    detected   \liviii\,  ~ $\lambda\lambda$670.8 nm   in spectra of V5668 Sgr  taken on day 7.
  The non detection of \livii\ ~  in our  observations implies  that the ejected gas has a temperature high enough that almost all  Li and Ca  atoms are ionised.     Normally Li is not detected in Novae outburst spectra and the   unique  \lii\,  detection  by \citet{izz15}
 implies  that the physical conditions in the ejecta     permit 
 the survival of  neutral  \liviii\ only in the very early stages. Since  \beviiii ~     decays to   \liviiii\      the non detection of   \liviii\  in our epochs requires that   \livii\  remains singly ionised while some  Na\,{\sc i} survives.

Since     \bevii\ $=$\livii\, the   $X(\mbox{\bevii})/X(\mbox{Ca})$  fraction derived here      corresponds to 
a \livii ~ logarithmic overabundances of   + 4.7 dex  
 with respect to the  meteoritic value \citep{lod09}, which is  even higher than  the overabundance of  4 dex    obtained by \cite{izz15}
 in Nova Centauri 2013.  Theoretically the amount of \livii\  is a sensitive function of the conditions achieved in the outburst and from the initial \iiihe\ of the companion star, and are expected to vary.  
 The  Novae in which \livii\  or \bevii\  have been detected to date are all {\it slow}  Novae characterised by  $ t_2 > 60^d $.
 However, it is quite remarkable that  large \bevii ~ yields are observed in all three Novae where   \beviiii ~ has been searched for. 
 For a total ejected mass of $\approx$ 10$^{-5} M{\sun}$    the observed  overproduction factor  of V5668 Sgr  implies  a production of    $M_{Li}$  $\approx$  $7 \cdot$ 10$^{-9} M{\sun}$.
 The global Nova 
rate in the  Galaxy is known within a factor two   
 ($25-50$ yr$^{-1}$)
 \citep{del94,sha16}.
and    the   {\it slow}  Novae  account for 
$\approx$ 10\% of the whole population \citep{del93}.  However, a  rate of 2   yr$^{-1}$   of {\it slow}  Nova events with the observed \livii\ overproduction  in a  Galaxy lifetime  of $\approx$ 10$^{10} $ yr   is enough to  produce $M_{Li}$ $\approx$ 140  $M{\sun}$. This is  comparable with the $M_{Li}$ $\approx$ 150   $M{\sun}$ ~  estimated to be present in the Milky Way inclusive of  the $M_{Li}$ $\approx$ 40 $M{\sun}$   produced in the  Big Bang  \citep{fie14}.    Thus,  the  {\it slow} Novae   could indeed be the main factories of \livii ~ in the Galaxy.

\section{Summary and conclusions}

 We have analysed  UVES high resolution observations of V5668 covering six outburst phases  from day 58 to day 89  from maximum. The evolution of the absorption offers   clear evidence in support of  the identification of \beviiii ~ by \citet{taj16}.  In particular,  the weakening of the \beviiii\ absorptions at a late epoch  shows that the iron-peak  species  are a minor  contaminant.  By means of   unsaturated   \beii ~  components  we derived an     abundance of   $N(\mbox{\bevii})/N(\mbox{Ca})$  $\approx$ 53-69 when the   $^{7}$Be  decay is taken into account. Assuming all the \bevii~ goes into \livii~ this corresponds to a \livii~ overproduction of 4.7 -  4.9 dex over the solar-meteoritic value. We then argue that a  rate of 2   yr$^{-1}$   of such  events in a  Galaxy lifetime, i.e.    only a small fraction of  all Novae,  could be responsible for  the production of the whole \livii~  required  from  {\it stellar}  sources.

We also notice that such a high \bevii\   production should increase the probability of detecting the 478-keV $\gamma$-ray photons emitted in the \bevii\  to \livii\  reaction which have been so far elusive despite several $\gamma$-ray searches.
 \section*{Acknowledgments}

This is an ESO DDT program and we  acknowledge   the ESO director for this opportunity and the ESO staff  for  care and competence in making the observations.
LI acknowledges support from the Spanish research project AYA 2014-58381-P.

 \bibliography{biblio_nova}

\begin{thebibliography}{20}
\expandafter\ifx\csname natexlab\endcsname\relax\def\natexlab#1{#1}\fi

\bibitem[{{Arnould} \& {Norgaard}(1975)}]{arn75}
{Arnould} M., {Norgaard} H., 1975, \aap, 42, 55

\bibitem[{{Boffin} {et~al}\mbox{.}(1993){Boffin}, {Paulus}, {Arnould}, \&
  {Mowlavi}}]{bof93}
{Boffin} H.~M.~J., {Paulus} G., {Arnould} M., {Mowlavi} N., 1993, \aap, 279,
  173

\bibitem[{{Cameron} \& {Fowler}(1971)}]{cam71}
{Cameron} A.~G.~W., {Fowler} W.~A., 1971, \apj, 164, 111

\bibitem[{{della Valle} \& {Duerbeck}(1993)}]{del93}
{della Valle} M., {Duerbeck} H.~W., 1993, \aap, 271, 175

\bibitem[{{della Valle} \& {Livio}(1994)}]{del94}
{della Valle} M., {Livio} M., 1994, \aap, 286

\bibitem[{{Denissenkov} {et~al}\mbox{.}(2013){Denissenkov}, {Herwig},
  {Bildsten}, \& {Paxton}}]{den13}
{Denissenkov} P.~A., {Herwig} F., {Bildsten} L., {Paxton} B., 2013, \apj, 762,
  8

\bibitem[{{Fields}, {Molaro} \& {Sarkar}(2014){Fields}, {Molaro}, \&
  {Sarkar}}]{fie14}
{Fields} B.~D., {Molaro} P., {Sarkar} S., 2014, ArXiv e-prints

\bibitem[{{Hernanz} {et~al}\mbox{.}(1996){Hernanz}, {Jose}, {Coc}, \&
  {Isern}}]{her96}
{Hernanz} M., {Jose} J., {Coc} A., {Isern} J., 1996, \apjl, 465, L27

\bibitem[{{Izzo} {et~al}\mbox{.}(2015){Izzo}, {Della Valle}, {Mason},
  {Matteucci}, {Romano}, {Pasquini}, {Vanzi}, {Jordan}, {Fernandez}, {Bluhm},
  {Brahm}, {Espinoza}, \& {Williams}}]{izz15}
{Izzo} L. {et~al.}, 2015, \apjl, 808, L14

\bibitem[{{Izzo} \& {et al}(2016)}]{izz16}
{Izzo} L., {et al}, 2016, in preparation

\bibitem[{{Jos{\'e}} \& {Hernanz}(1998)}]{jos98}
{Jos{\'e}} J., {Hernanz} M., 1998, \apj, 494, 680

\bibitem[{{Lodders}, {Palme} \& {Gail}(2009){Lodders}, {Palme}, \&
  {Gail}}]{lod09}
{Lodders} K., {Palme} H., {Gail} H.-P., 2009, Landolt B{\"o}rnstein

\bibitem[{{Rebolo}, {Beckman} \& {Molaro}(1988){Rebolo}, {Beckman}, \&
  {Molaro}}]{reb88}
{Rebolo} R., {Beckman} J.~E., {Molaro} P., 1988, \aap, 192, 192

\bibitem[{{Romano} {et~al}\mbox{.}(1999){Romano}, {Matteucci}, {Molaro}, \&
  {Bonifacio}}]{rom99}
{Romano} D., {Matteucci} F., {Molaro} P., {Bonifacio} P., 1999, \aap, 352, 117

\bibitem[{{Seach}(2015)}]{sea15}
{Seach} J., 2015, Central Bureau Electronic Telegrams, 4080

\bibitem[{{Shafter}(2016)}]{sha16}
{Shafter} A.~W., 2016, ArXiv e-prints

\bibitem[{{Spitzer}(1998)}]{spi98}
{Spitzer} L., 1998, {Physical Processes in the Interstellar Medium}. p. 335

\bibitem[{{Starrfield} {et~al}\mbox{.}(1978){Starrfield}, {Truran}, {Sparks},
  \& {Arnould}}]{sta78}
{Starrfield} S., {Truran} J.~W., {Sparks} W.~M., {Arnould} M., 1978, \apj, 222,
  600

\bibitem[{{Tajitsu} {et~al}\mbox{.}(2015){Tajitsu}, {Sadakane}, {Naito},
  {Arai}, \& {Aoki}}]{taj15}
{Tajitsu} A., {Sadakane} K., {Naito} H., {Arai} A., {Aoki} W., 2015, \nat, 518,
  381

\bibitem[{{Tajitsu} {et~al}\mbox{.}(2016){Tajitsu}, {Sadakane}, {Naito},
  {Arai}, {Kawakita}, \& {Aoki}}]{taj16}
{Tajitsu} A., {Sadakane} K., {Naito} H., {Arai} A., {Kawakita} H., {Aoki} W.,
  2016, \apj, 818, 191

\end{thebibliography}

\end{document}